\documentclass[11pt]{article}
\usepackage{moriond_proc,epsfig}

\def\Journal#1#2#3#4{{#1} {\bf #2}, #3 (#4)}

\def\NPB{{\em Nucl. Phys.} B}
\def\PLB{{\em Phys. Lett.}  B}
\def\PRL{\em Phys. Rev. Lett.}
\def\PRD{{\em Phys. Rev.} D}

\def\be{\begin{equation}}
\def\ee{\end{equation}}
\def\bea{\begin{eqnarray}}
\def\eea{\end{eqnarray}}

\newcommand{\xilg} {\Xi \rightarrow \Lambda \gamma}
\newcommand{\klpipienu} {K_L \rightarrow \pi^0 \pi^{\pm} e^{\mp} \nu}
\newcommand{\kopienug} {K^0 \rightarrow \pi^{\pm} e^{\mp} \nu \gamma}
\newcommand{\kopienu} {K^0 \rightarrow \pi^{\pm} e^{\mp} \nu}
\newcommand{\kspimumu} {K_S \rightarrow \pi^0 \mu^+ \mu^-}
\newcommand{\klpipipic} {K_L \rightarrow \pi^0 \pi^+ \pi^-}
\newcommand{\klmumugg} {K_L \rightarrow \mu^+ \mu^- \gamma \gamma}
\newcommand{\kspiee} {K_S \rightarrow \pi^0 e^+ e^-}
\newcommand{\klpiee} {K_L \rightarrow \pi^0 e^+ e^-}
\newcommand{\klpimumu} {K_S \rightarrow \pi^0 \mu^+ \mu^-}
\newcommand{\kspill} {K_S \rightarrow \pi^0 l^+ l^-}
\newcommand{\klpill} {K_L \rightarrow \pi^0 l^+ l^-}
\newcommand{\klpinunubar} {K_L \rightarrow \pi^0 \nu \bar{\nu}}
\newcommand{\klpigg} {K_L \rightarrow \pi^0 \gamma \gamma}
\newcommand{\dalitz} {\pi^0 \rightarrow ee \gamma}
\newcommand{\Xilamppipio} {\Xi^{0}\rightarrow\Lambda (p \pi^{-})\pi^0 }
\newcommand{\Xilampevpio} {\Xi^{0}\rightarrow\Lambda (p e^{-} \nu)\pi^0 }

\newcommand{\kleth} {K_L \rightarrow \pi^{\pm} e^{\mp} \nu }
\newcommand{\kspiopio} {K_S \rightarrow \pi^0 \pi^0 }
\newcommand{\kleegg} {K_L \rightarrow ee \gamma \gamma }
\newcommand{\klpienug} {K_L \rightarrow \pi^\pm e^\mp \nu \gamma}
\newcommand{\Xisigmappioenv} {\Xi^0 \rightarrow \Sigma^{+}(p \pi^0) e^{-} \nu}

\newcommand{\kspipic} {K_S \rightarrow  \pi^+ \pi^-}

\begin{document}
\vspace*{4cm}
\title{Latest Results from NA48 and NA48/1}

\author{ M. W. SLATER }

\address{Department of Physics, High Energy Physics, Cavendish Laboratory, \\
Madingley Road, Cambridge CB3 0HE, England}

\maketitle\abstracts{
The first observations of the rare decays $\kspiee$ and $\kspimumu$
have been made by the NA48/1 collaboration at the CERN SPS accelerator.
From high intensity $K_S$ data collected during the 2002 run, clean signals
of 7 $\kspiee$ events and 6 $\kspimumu$ events were observed, giving branching ratio 
measurements of 
$BR(\kspiee) = 5.8^{+2.9}_{-2.4} \times 10^{-9}$ and $BR(\kspimumu) =
2.9^{+1.5}_{-1.2} \times 10^{-9}$. 
These results constrain the indirect CP violating component of the
corresponding $K_L$ decays. Other recent results from NA48 are also
presented.}

\section{Introduction}

The NA48 experiment was originally designed to measure the CP violation parameter
Re($\frac{\epsilon '}{\epsilon}$) via a high statistics comparison of $K_S \rightarrow \pi \pi$
and $K_L \rightarrow \pi \pi$ decays. A beamline consisting
of simultaneous $K_S$ and $K_L$ beams was used, with a tagging system allowing discrimination between
the two types of kaon. Rare $K_L$ decay and precision
measurements could be carried out in parallel. In 2002, for the NA48/1 phase of the experiment,
the $K_L$ beam was removed and the proton intensity on the $K_S$ 
target was increased by a factor $\sim 1000$, allowing very rare $K_S$ and hyperon
decay searches to be pursued.

The kaon beams were produced using 400 GeV$/c$ protons provided 
by the Super Proton Synchrotron (SPS) at CERN,
incident on two separate beryllium targets. The NA48 detector
consisted of the following principal sub-detectors: a magnetic spectrometer consisting 
of 4 drift chambers separated by a dipole magnet, 
a high resolution liquid krypton electromagnetic calorimeter, 
an iron scintillator hadronic calorimeter 
and a muon system consisting of 3 planes of scintillator shielded by 80cm thick iron walls.\cite{na48}

\section{Physics Motivation for {\boldmath $\kspill$} Searches}


The decay mode $\klpinunubar$ is direct CP violating and can
be used to determine the parameter $\eta$ in the Wolfenstein parameterisation of the CKM matrix.
This channel is theoretically very clean but is experimentally challenging 
due to the missing energy from the neutrinos. A second possibility to measure 
$\eta$ is from the decay $\klpill$. This decay is
easier to measure experimentally as all the decay products are detectable, but is
 more complicated
theoretically as it has contributions from CP conserving and both direct and indirect
CP violating components, which also interfere. The CP conserving component can be predicted from
a measurement of $\klpigg$ while the indirect CP violating component can be 
predicted from a measurement of
$\kspill$. Any measurement of $BR(\klpill)$, and consequently, $\eta$,
 requires the measurement of $BR(\kspill)$ to disentangle the direct from the indirect
CP violating components.

\section{Analysis Strategy for {\boldmath $\kspiee$} and {\boldmath $\kspimumu$}}

Both $\kspiee$ and $\kspimumu$ were predicted to have very small branching ratios,
($\sim10^{-9}$) and so, even with a flux of $\sim 3 \times 10^{10}$ $K_S$ decays,
only a handful of events were expected from either channel. Therefore, to avoid biasing
the results, a blind analysis procedure was employed. A
$2.5\sigma_{m_K} \times  2.5\sigma_{\pi^0}$ signal region and a 
$6.0\sigma_{m_K} \times  6.0\sigma_{\pi^0}$ control region were defined.
 Both regions were kept masked during the
background studies. Only after the background contributions to the signal had been estimated and
the cuts fixed was the control region unmasked. Final selection changes could then
be made if required before the signal region was unmasked.

\begin{figure}[t]
\begin{center}
\epsfig{file=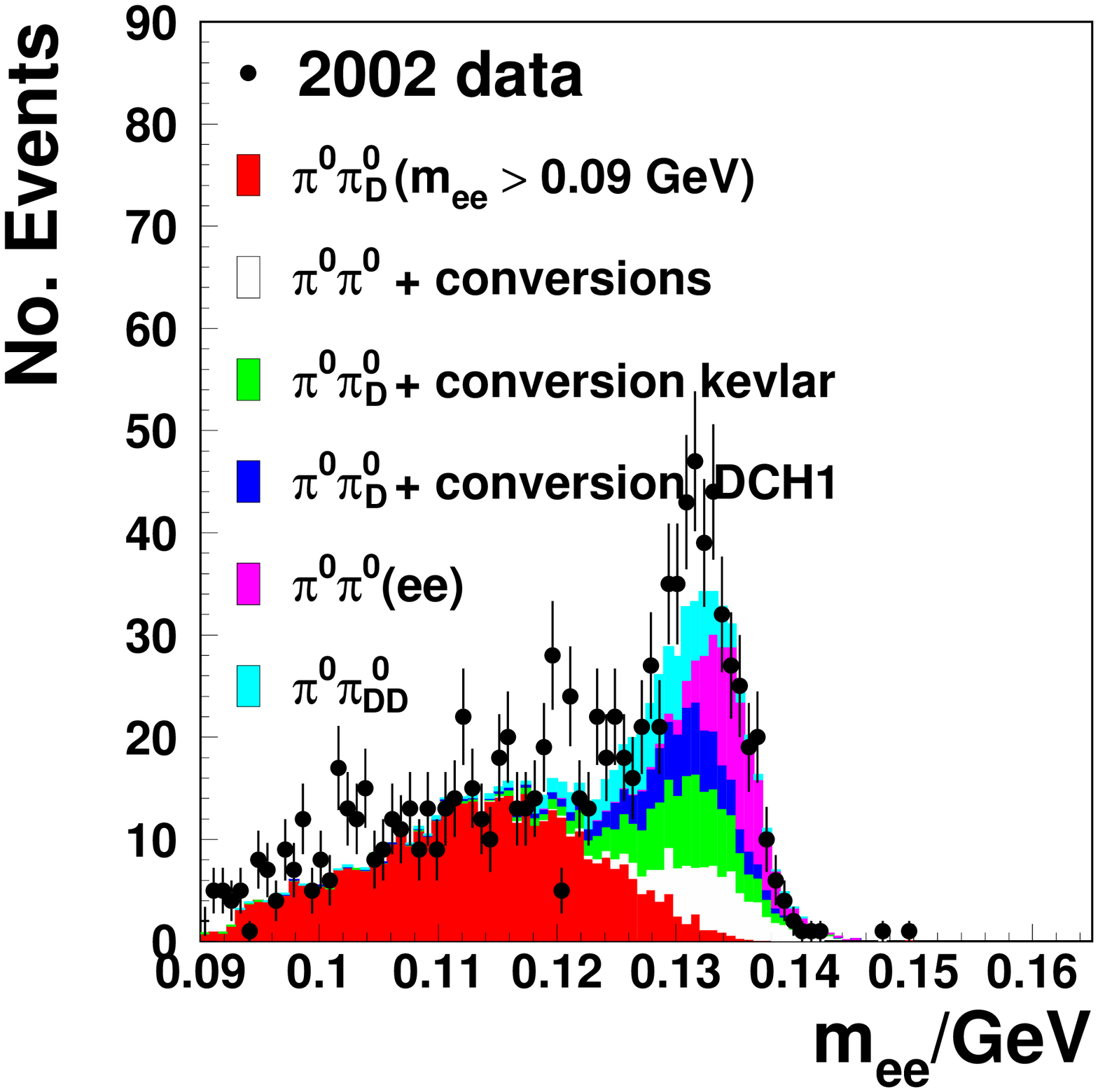,width=60mm}
\hspace{0.5cm}
\epsfig{file=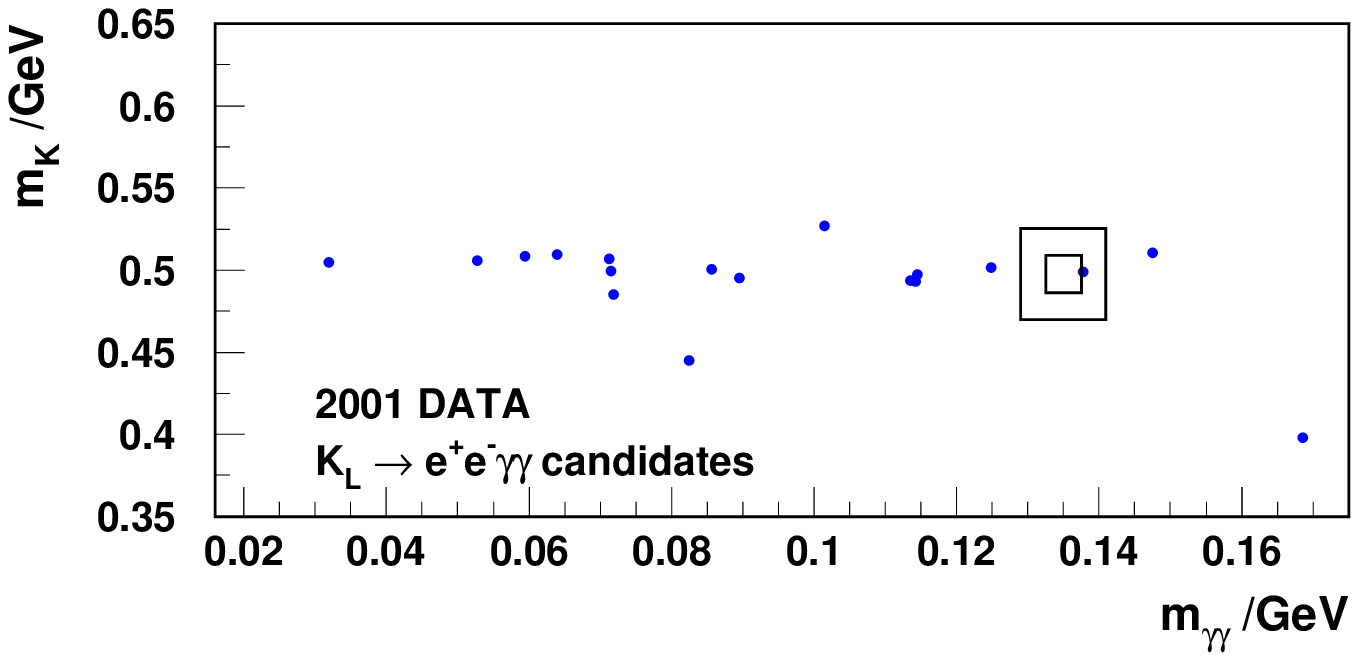,width=90mm}
\caption{\label{piee_back} (a) $m_{ee}$ distribution of the Dalitz decay and conversion background.
Reasonable agreement between data (points) and Monte Carlo (curves) can be seen.
(b) 2001 data showing the distribution of the $\kleegg$ background across the signal region.}
\end{center}
\end{figure}

\section{{\boldmath $\kspiee$}}

\subsection{Signal Selection}

$\kspiee$ candidates were selected with $40 < E_K < 240 \mathrm{GeV}$ and within
2.5 $K_S$ lifetimes of the end of the final collimator. Two oppositely charged tracks, with $E/p > 0.95$
 and forming a good vertex, were required, as well as two clusters in the calorimeter that were not
associated to tracks. 

\subsection{Background}

Backgrounds for the $\kspiee$ decay were divided into two categories: physical
background (arising from a single kaon decay), and accidental background (arising
from two separate, overlapping kaon decays).
Three significant sources of physical background were identified. The first came
from Dalitz decays ($\dalitz$) and photon conversion in the various parts of the
detector. These backgrounds were studied extensively using Monte Carlo simulation (see fig.~\ref{piee_back}a).
Reasonable agreement with data was found in the $m_{ee}$ distribution around the $\pi^0$ mass.
A conservative cut of $m_{ee} > 0.165 \,\mathrm{GeV}/c^2$ was applied to remove this background,
with a corresponding $48\%$ loss in acceptance.

The second significant physical background was identified as $\kleegg$. 
This background spread across
the $m_{\gamma \gamma}$ signal region as can be seen in fig.~\ref{piee_back}b. 
Using $K_L$ data taken in 2001, the background was estimated to be $0.08^{+0.03}_{-0.02}$ events.

\begin{figure}[t]
\begin{center}
\epsfig{file=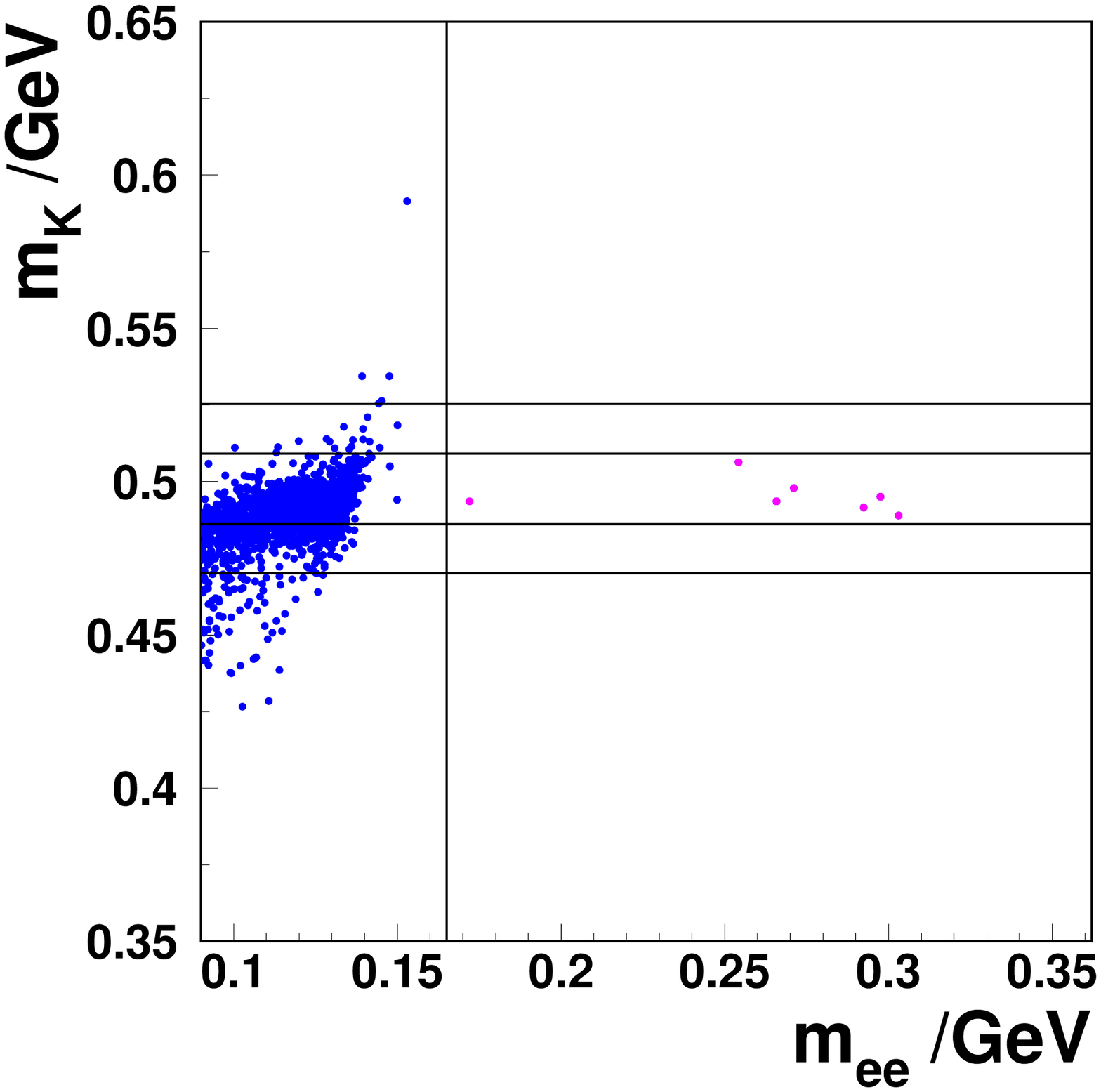,width=70mm}
\hspace{1cm}
\epsfig{file=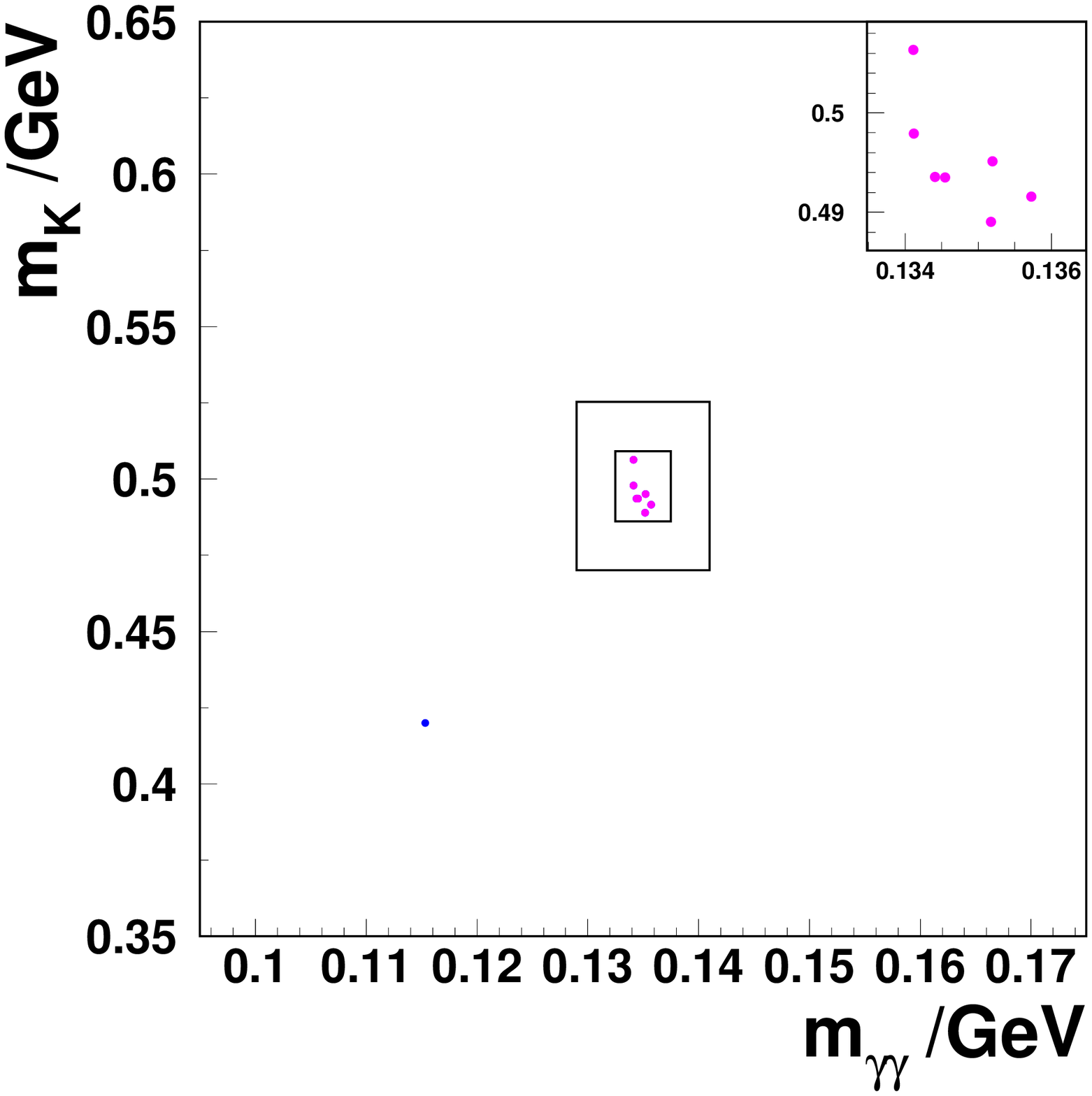,width=70mm}
\caption{\label{piee} The (a) $m_{\gamma\gamma}$ vs. $m_{ee}$ and
(b) $m_{\gamma\gamma}$ vs. $m_K$ mass planes showing $\kspiee$ candidate events. 
The large background from dalitz decays and photon conversions can be seen at low $m_{ee}$}
\end{center}
\end{figure}

The final physical background came from various hyperon decay channels. The beam contained a significant ($\sim 10^9$)
flux of neutral hyperons and the channels $\Xilamppipio$, $\Xilampevpio$ and $\Xisigmappioenv$ were identified as possible
background sources. A cut on the momentum asymmetry of the decay products reduced these backgrounds to a negligible level.

The accidental background was dominated by overlapping fragments of two decays, e.g
($\kleth$) + ($\kspiopio$). This background was suppressed by cutting on the spread of
track and cluster times ($|\Delta t| < 3 \mathrm{ns}$).
To estimate the remaining accidental background, the timing cuts were relaxed and the
time sidebands investigated. No events were found in the out of time
($3 \mathrm{ns} < |\Delta t| < 50 \mathrm{ns}$)
signal region. Therefore, an extrapolation was made from the out of time control
region to the out of time signal region, taking into account the shape of the
background from Monte Carlo simulation. A further time extrapolation
could then be performed to get the final background estimate for the in time region
($|\Delta t| < 3 \mathrm{ns}$) of 0.069 events.

\subsection{Result}

The overall background estimate was $0.15^{+0.05}_{-0.04}$ events.
The control region was unmasked and no events were found. No changes to the selection were needed
before the signal region was unmasked. A signal of 7 events was found,
giving the result: \cite{piee_paper}
\begin{equation}
BR(\kspiee) = {(5.8^{+2.8}_{-2.3}(stat)\pm0.8(syst)) \times 10^{-9}}
\end{equation}

\noindent where the systematic error is dominated by 
uncertainties arising from the extrapolation 
to the full $m_{ee}$ region. The candidate events are shown in fig.~\ref{piee}.

\section{{\boldmath $\kspimumu$}}

\begin{figure}[t]
\hspace{0.05\textwidth}
\epsfig{file=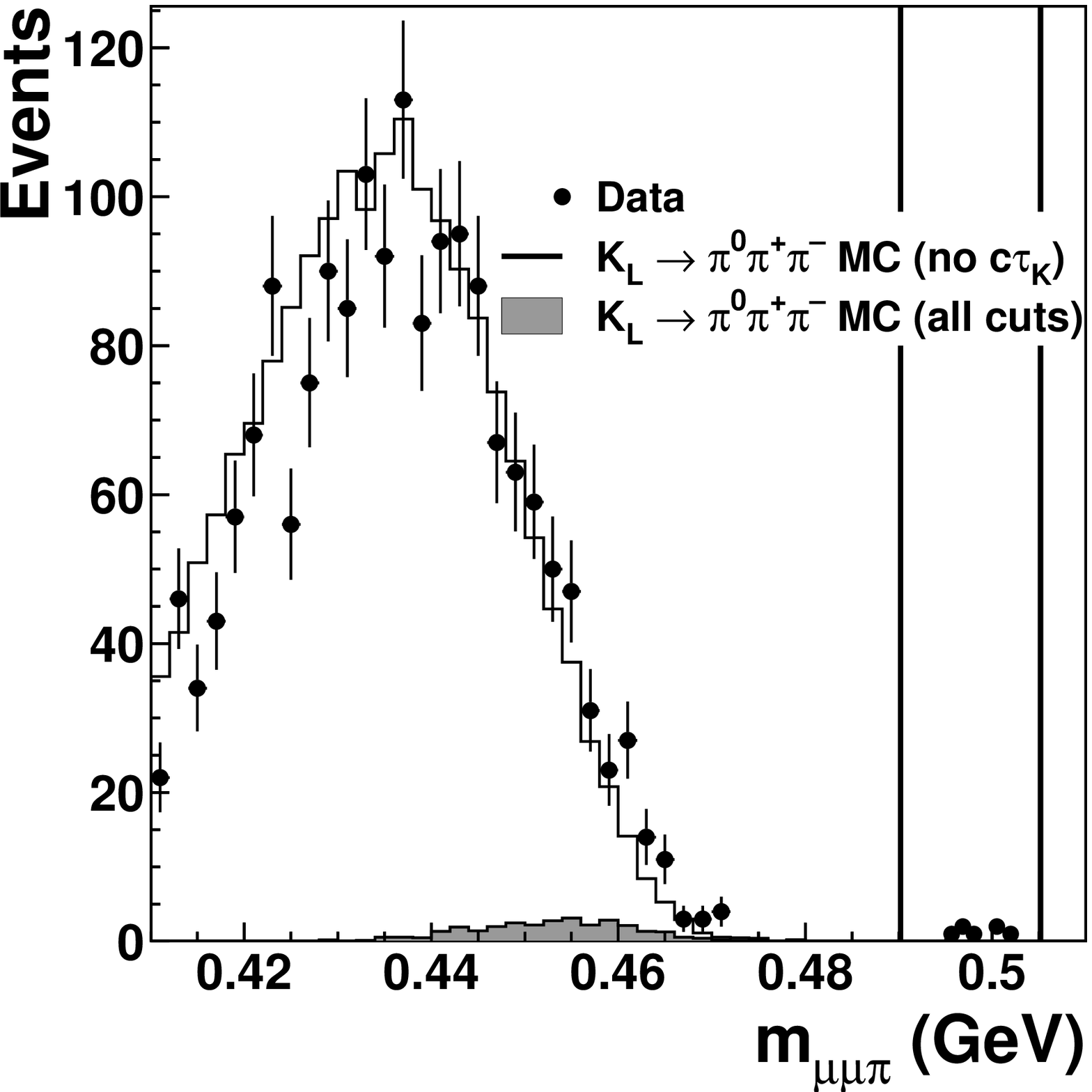,width=60mm}

\vspace*{-6.5cm}
\hspace{0.45\textwidth}
\epsfig{file=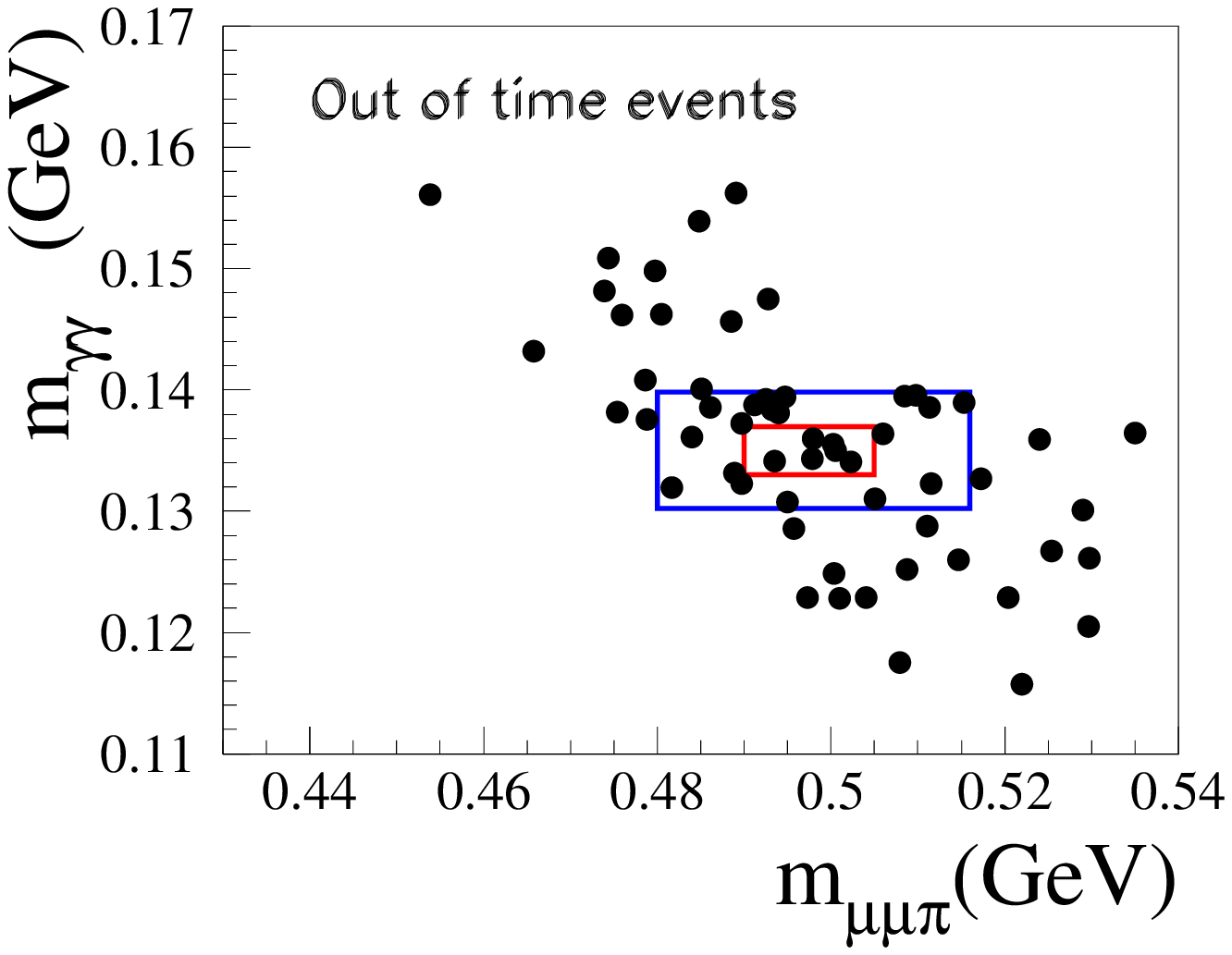,width=80mm}

\vspace*{-2cm}
\caption{\label{piuu_back} (a) The kaon mass region with the $\klpipipic$ background. Reasonable agreement
between data and monte carlo (24 times the data) is shown.
(b) Distribution of out of time events in the $m_K$ vs. $m_{\gamma\gamma}$
plane.}
\end{figure}

\subsection {Signal Selection}

$\kspimumu$ candidates were selected with $60 < E_K < 200 \mathrm{GeV}$ and within
3 $K_S$ lifetimes of the end of the final collimator. Two oppositely charged tracks were
required that formed a good vertex, as well as two clusters in the calorimeter that were not
associated to tracks. The tracks had to leave minimum energy in the calorimetry and have
associated hits in the muon detector.

\subsection{Background}

As with $\kspiee$, the backgrounds were divided into physical and accidental sources.
Three dominant sources of physical background were identified. The first was
$\klpipipic$ where the two charged pions had decayed in flight. This background was studied
extensively in Monte Carlo, (see fig.~\ref{piuu_back}a) and estimated to be $\leq 0.019$
events. The second significant background was $\klmumugg$. This, again, was studied with
Monte Carlo and estimated to contribute $0.04 \pm 0.04$ events to the overall background.
The last physical background came from hyperon decays. As with $\kspiee$, a cut on the momentum
asymmetry of the decay products was used to reduce this background to a negligible level.

The accidental background was treated similarly to $\kspiee$, by loosening the
timing cuts and extrapolating into the in time ($| \Delta t| < 1.5$ns) region. 6 events were
found in the out of time ($-115 \mathrm{ns} < \Delta t < 60 \mathrm{ns}$) signal region, leading to a background 
estimate of $0.18^{+0.18}_{-0.11}$ events. The distribution of these events in the mass plane
is shown in fig.~\ref{piuu_back}b.

\subsection{Result}

\begin{figure}[t]
\begin{center}
\epsfig{file=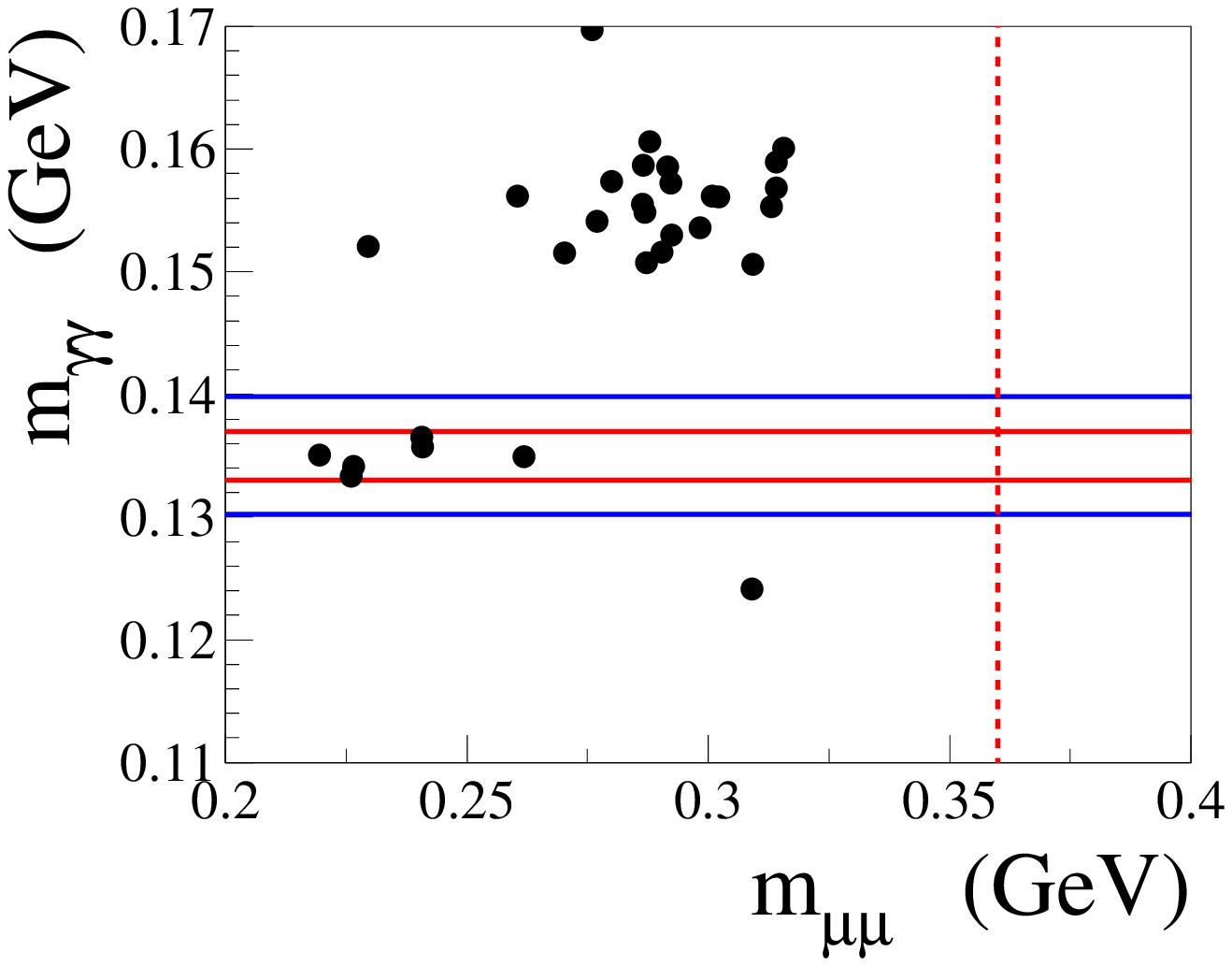,width=75mm}
\hspace{0.5cm}
\epsfig{file=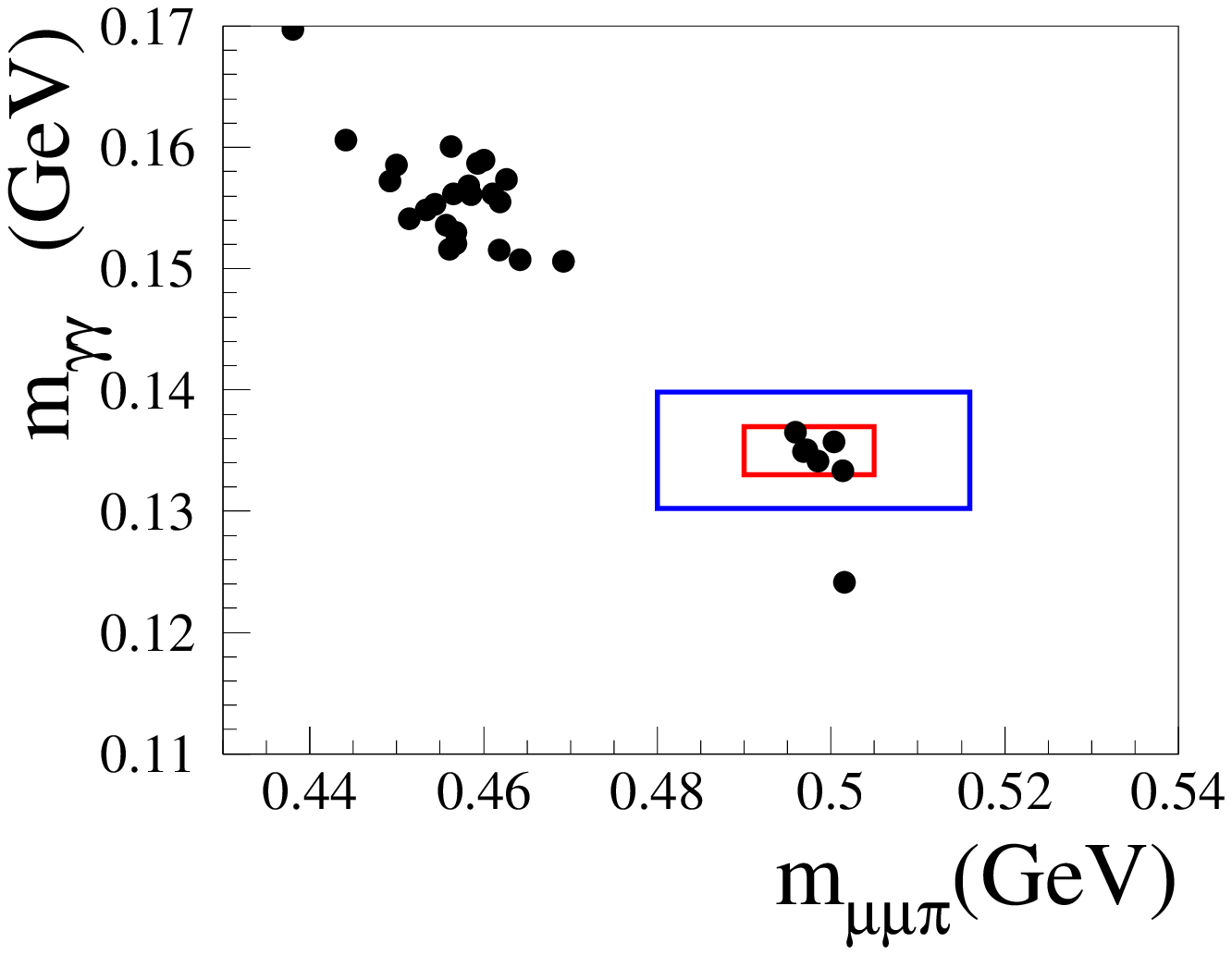,width=75mm}
\vspace{-2cm}
\caption{\label{piuu} The (a) $m_{\gamma\gamma}$ vs. $m_{ee}$ and
(b) $m_{\gamma\gamma}$ vs. $m_K$ mass planes showing $\kspimumu$ candidate events. The events at 
large $m_{\gamma \gamma}$ are consistent with $\klpipipic$ background and the single event at low $m_{\gamma\gamma}$ is
consistent with accidental background}
\end{center}
\end{figure}

The overall background estimate was $0.22^{+0.19}_{-0.12}$ events.
The control region was unmasked and no events were found. No changes to the selection were needed,
the signal region was unmasked and 6 events were found. 
This led to the result:
\begin{equation}
BR(\kspimumu) = {(2.9^{+1.5}_{-1.2}(stat)\pm0.2(syst)) \times 10^{-9}}
\end{equation}
\noindent where the systematic error comes from uncertainties in the normalisation,
obtained from $\kspipic$ decays.
The candidate events are shown in fig.~\ref{piuu}.

\section{Interpretation of the {\boldmath $\kspill$} Measurements}

The form factor for $\kspill$ decays can be described in Chiral Perturbation Theory by a
first order polynomial:
\begin{equation}
 W(z) \simeq G_Fm^2_K (a_S + b_S z)
\end{equation}

\noindent where $z = m^2_{ll}/m^2_K$. This leads to the following predictions for the $\kspill$ branching ratios:\cite{pill_paper}
\begin{equation}
BR(\kspiee) = [0.01-0.76a_S-0.21b_S+46.5a^2_S+12.9a_Sb_S+1.44b^2_S]\times10^{-10}
\end{equation}
\begin{equation}
BR(\kspimumu) = [0.07-4.52a_S-1.50b_S+98.7a^2_S+57.7a_Sb_S+8.95b^2_S]\times10^{-11}
\end{equation}

\noindent The Vector Meson Dominance (VMD) model predicts the ratio $b_S / a_S = 0.4$, allowing
$|a_S|$ to be extracted for both channels:
\begin{equation}
BR(\kspiee)\simeq5.2\times 10^{-9} a^2_S  \hspace*{0.3cm}  
\Rightarrow |a_S|_{\pi^0 e e} = 1.06^{+0.26}_{-0.21}\pm0.07
\end{equation}
\begin{equation}
BR(\kspimumu)\simeq1.2\times 10^{-9} a^2_S  \hspace*{0.3cm}  
\Rightarrow |a_S|_{\pi^0 \mu \mu} = 1.55^{+0.38}_{-0.32}\pm0.05
\end{equation}

\noindent These results for $|a_S|$ agree within errors.

\begin{figure}[t]
\begin{center}
\epsfig{file=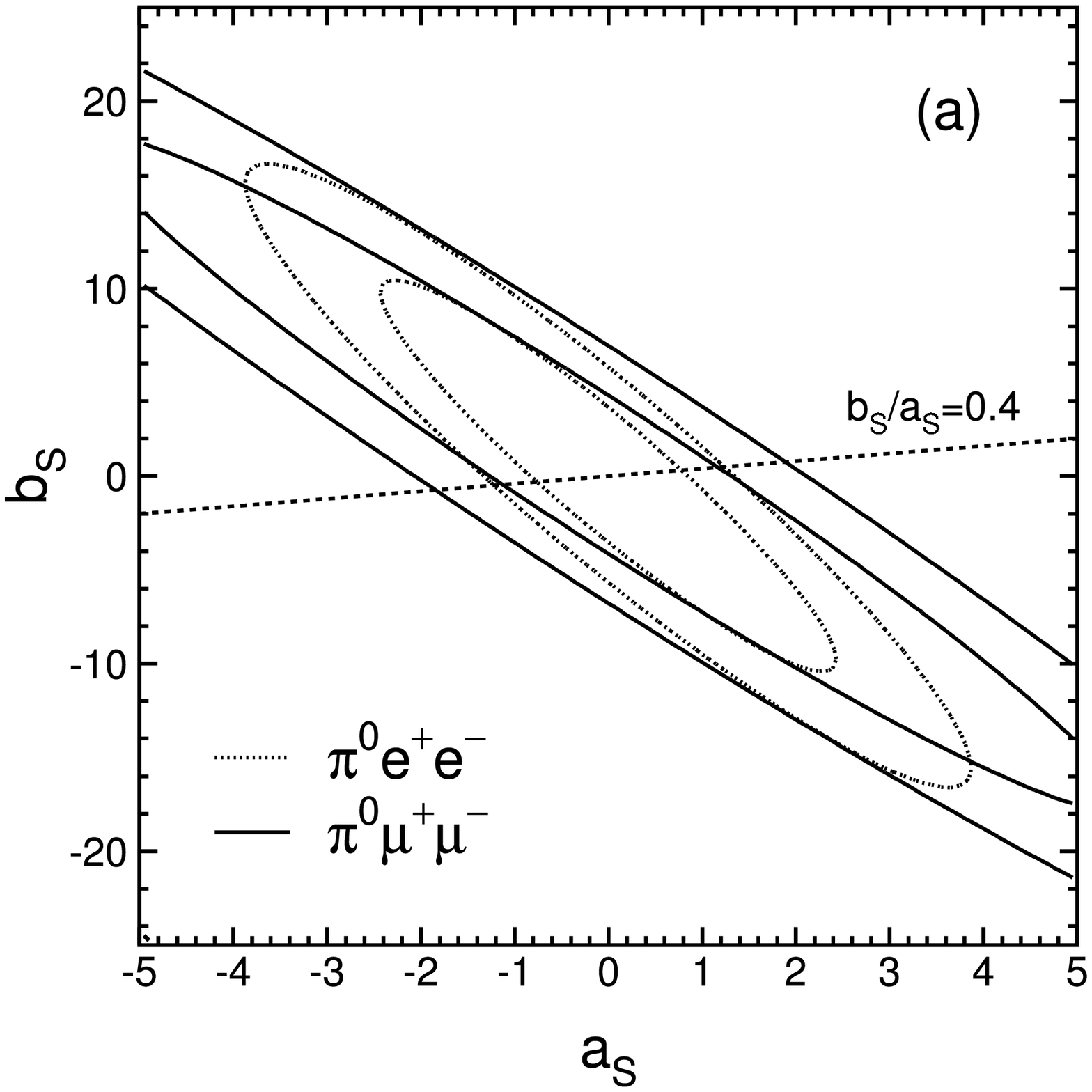,width=60mm}
\hspace{1cm}
\epsfig{file=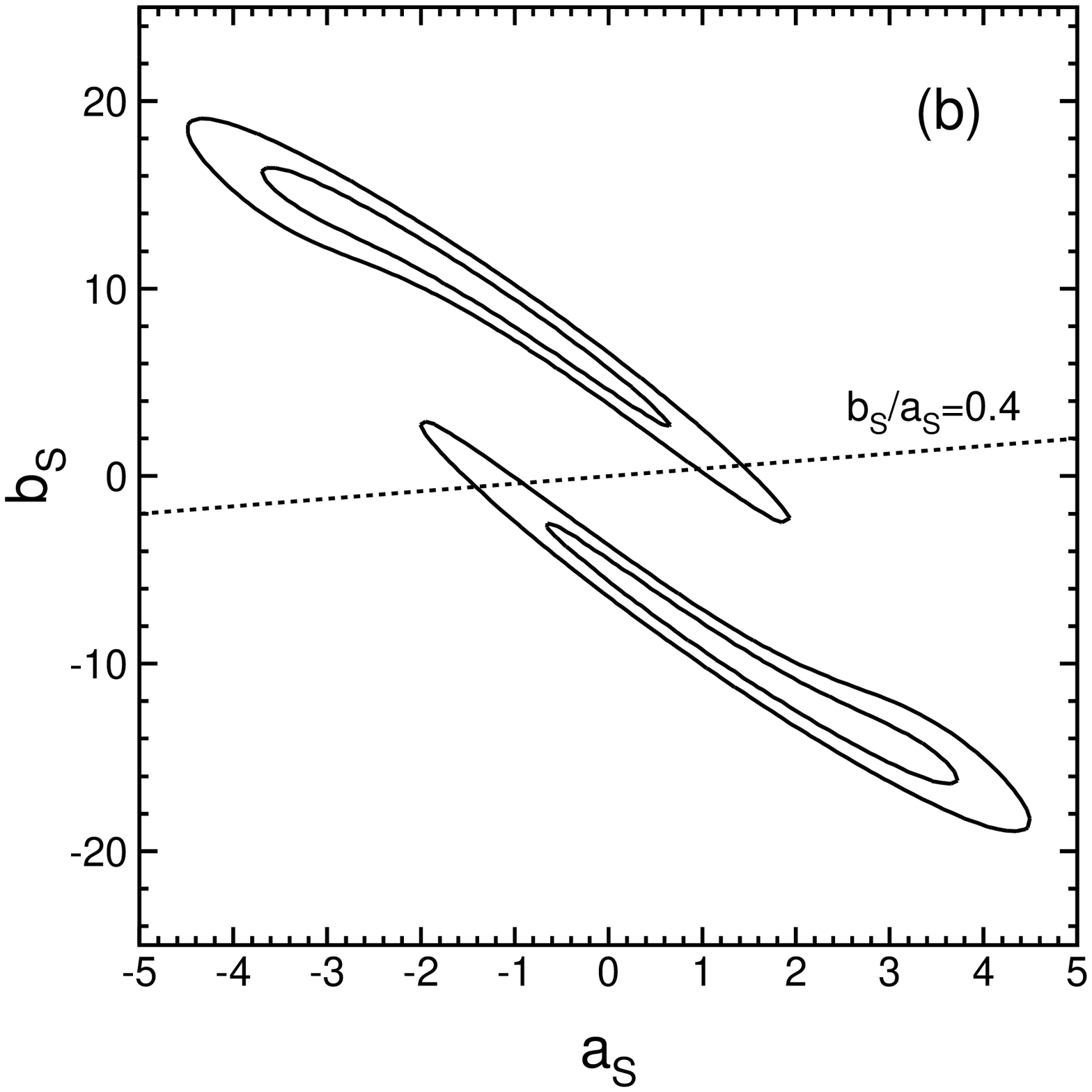,width=60mm}
\caption{\label{asbs} (a) Allowed regions of $a_S$ and $b_S$ determined from the observed number
of $\kspiee$ and $\kspimumu$ events separately. The region between the inner and outer 
elliptical contours is the allowed region at $68\%$ CL. (b) Allowed regions of $a_S$ and $b_S$ for
the $\kspiee$ and $\kspimumu$ channels combined. The contours delimit the $1\sigma$ and $2\sigma$
allowed regions from the combined log-likelihood. The dashed straight line in both plots corresponds
to $b_S = 0.4 a_S$, as predicted by the VMD model.}

\end{center}
\end{figure}

By combining the $ee$ and $\mu\mu$ results in a log-likelihood fit, $a_S$ and $b_S$ can be determined
separately. 
As can be seen from fig.~\ref{asbs}, the observed $\kspiee$ and $\kspimumu$ rates
 are compatible with both each other and the VMD model.


\section{Implications for {\boldmath $\klpill$}}

The CP violating component of the $\klpill$ branching ratio is given by:
\begin{equation}
BR(\klpill)_{CPV} \times 10^{12} = C_{IND}\pm C_{MIX}\left(\frac{\mathrm{Im}(\lambda_t)}{10^{-4}}\right) 
\pm C_{DIR}\left(\frac{\mathrm{Im}(\lambda_t)}{10^{-4}}\right)^2
\end{equation}

\noindent where $C_{DIR}$ is the direct CPV component, $C_{IND} \sim BR(\kspill)$
is the indirect CPV component, $C_{MIX} \sim \sqrt{BR(\kspill)}$ is the interference term
and $\mathrm{Im}(\lambda_t) = \mathrm{Im}(V^*_{ts} V_{td})$.

The indirect CPV component and the interference term depend on the
parameter $|a_S|$. Using the measured values of $|a_S|$ given above and taking \cite{lamt}
$\mathrm{Im}(\lambda_t) = (1.36\pm0.12)\times 10^{-4}$ gives the following predictions for the 
central value of the $\klpill$
branching ratios:
\begin{equation}
BR(\klpiee)_{CPV} \times 10^{12} \simeq 17_{\mathrm{indirect}} \pm 9_{\mathrm{interference}}
 + 5_{\mathrm{direct}}
\end{equation}
\begin{equation}
BR(\klpimumu)_{CPV} \times 10^{12} \simeq 9_{\mathrm{indirect}} \pm 3_{\mathrm{interference}} + 1_{\mathrm{direct}}
\end{equation}

\noindent where the sign ambiguity reflects the uncertainty in the sign of $a_S$.

\section{{\boldmath $\klpienug$}}

The most precise published measurement of $BR(\kopienug)$ comes from the KTeV
collabora-
tion:\cite{ktev}
\begin{equation}
BR(\kopienug)/BR(\kopienu) = (0.908\pm0.008^{+0.013}_{-0.012})\%,
\end{equation}

\noindent which is significantly below current theroretical predictions. This branching 
ratio has now been measured by NA48 using data from 1999. In order to reduce
the uncertainties in the measurement due to radiative corrections, a
 model independent method was developed. The PHOTOS \cite{photos} package
was used for the initial Monte Carlo generation. The Monte Carlo events were then weighted
with the $\theta^*_{e\gamma}$ distribution, (the angle between the electron and the photon
in the kaon rest frame).
 This gave good agreement between data and Monte Carlo
for all distributions. The preliminary branching ratio measurement using
 this model independent method was found to be:
\begin{equation}
BR(\kopienug)/BR(\kopienu) = (0.964\pm0.008^{+0.012}_{-0.011})\%
\end{equation}

\begin{figure}[t]
\begin{center}
\epsfig{file=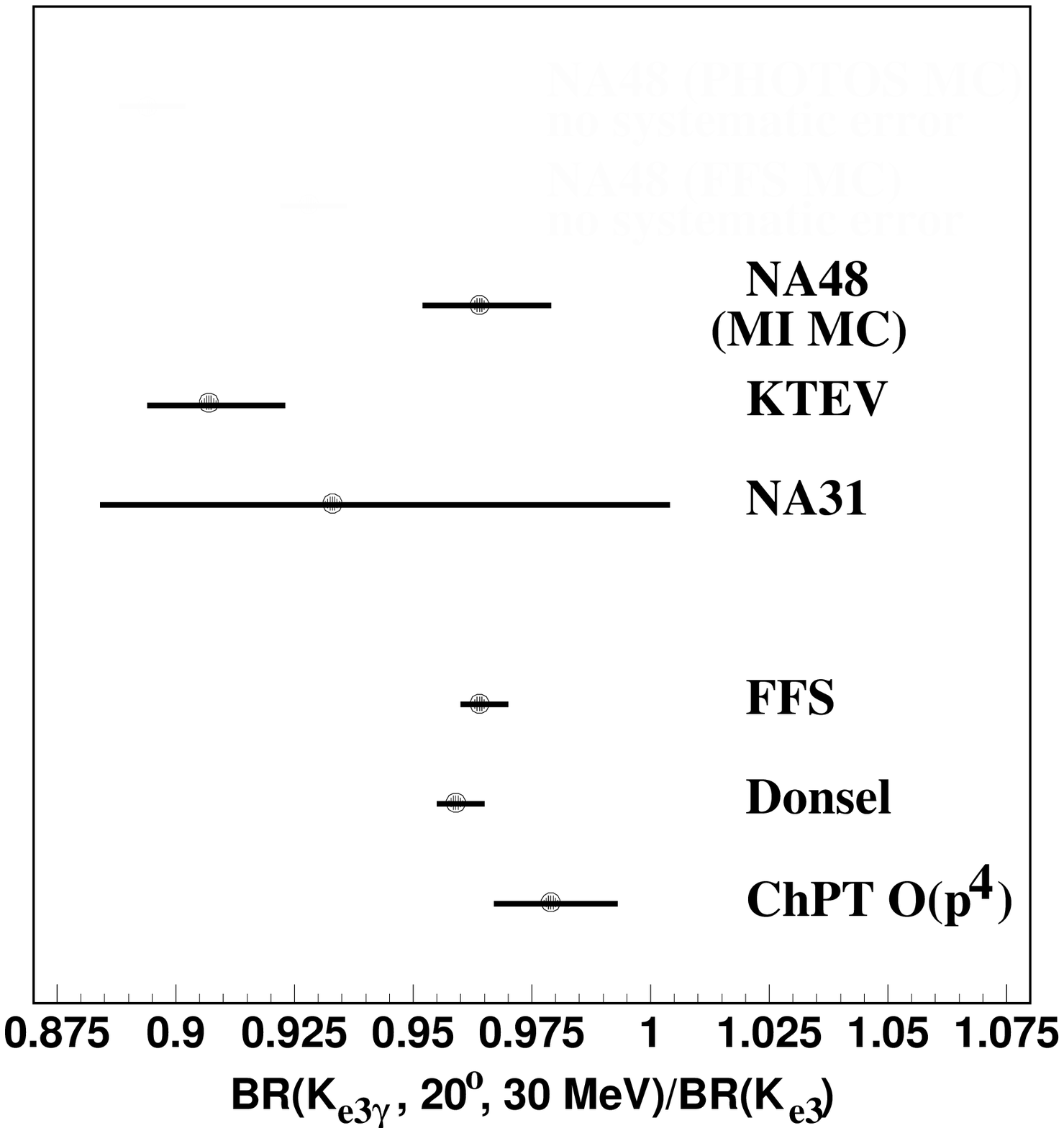,width=80mm}
\caption{\label{ketg} The measured value of $BR(\kopienug)/BR(\kopienu)$ compared
 with the published result from the KTeV collaboration and theoretical predictions. }
\end{center}
\end{figure}

\noindent This is in good agreement with recent theoretical predictions, (Fearing, Fischbach and 
Smith (FFS) \cite{ketg_paper1}$^,$\cite{ketg_paper2},
Doncel\cite{ketg_paper3} and Chiral Perturbation Theory calculations \cite{ketg_paper4}$^,$\cite{ketg_paper5}), as can be seen from fig.~\ref{ketg}. If the FFS method
is used to apply the radiative corrections, as in the KTeV analysis,
the result is found to be in agreement with the KTeV measurement.

\section{{\boldmath $\klpipienu$}}

A high precision measurement of the branching ratio and form factors for the 
$\klpipienu$ decay was made using $K_L$ data from 2001. This decay 
provides a good test of Chiral Perturbation Theory (CHPT) predictions for long distance meson 
interactions and the form factor measurements allow the determination of the CHPT parameter, $L_3$ .
\cite{kef_paper1}$^,$\cite{kef_paper2}

The main background to this decay was $\klpipipic$ where a pion
was misidentified as an electron. To minimise this background, a neural network was used to distinguish
between pions and electrons that took into account geometric characteristics
of the showers and tracks. This reduced the background to $1.13\%$.

A signal of 5464 events was found with a background of 62 events. This gave a preliminary branching ratio 
measurement of:
\begin{equation}
BR(\klpipienu) = (5.21\pm0.07_{stat}\pm0.09_{syst})\times10^{-5}
\end{equation}

\noindent Assuming a V-A structure for the matrix element, the form factor parameters were then measured to be:\\

\noindent \hspace*{5cm} $\bar{f_s} = 0.052\pm0.006_{stat}\pm0.002_{syst}$\\
\hspace*{5cm} $\bar{f_p} = -0.051\pm0.011_{stat}\pm0.005_{syst}$\\
\hspace*{4.95cm} $\lambda_g = 0.087\pm0.019_{stat}\pm0.006_{syst}$\\
\hspace*{5.1cm} $\bar{h} = -0.32\pm0.12_{stat}\pm0.07_{syst}$
\newpage
\noindent This led to the following value for $L_3$:
\begin{equation}
L_3 = (-4.1\pm0.2) \times 10 ^{-3}
\end{equation}

\section{{\boldmath $\xilg$}}

Using data from a short test run from 1999 with a high intensity $K_S$ beam alone,
 730 $\xilg$ events were found with a background of $58.2\pm7.8$ events. This led to a 
branching ratio measurement of:\cite{xi_paper}
\begin{equation}
BR(\xilg) = (1.16\pm0.05_{stat}\pm0.06_{syst})\times10^{-3}
\end{equation}

\noindent The statistics were large enough to enable a significant measurement of the decay
asymmetry to be made:
\begin{equation}
\alpha(\xilg) = -0.78\pm0.18_{stat}\pm0.06_{syst}
\end{equation}

\noindent This is the first evidence for a non-zero asymmetry in this decay mode. A much larger
sample from the 2002 run is currently under analysis.

\section*{Acknowledgments}

It is a pleasure to thank the technical staff of the participating laboratories, universities
and affiliated computing centres for their efforts in the construction of the NA48 detector
apparatus, in the operation of the experiment, and in the processing of the data.

\end{document}